\documentclass{jaa}
\usepackage{graphicx,url}

\begin{document}

%%paper title
%%For line breaks \\ can be used within title 
\title{Superfluidity and Superconductivity in Neutron Stars}

%%author names are separated by comma (,) 
%%use \and before the last author name 
%%use a * along with the number separated by comma
%% for the  author for correspondence
%%\textsuperscript{number} is used for affiliation
%%\affilOne, \affilTwo etc., upto \affilTwentyfive is possible
%%Please note the first letter after \affil is capitalised in the command
%%

\author{N. Chamel\textsuperscript{1,*}}
\affilOne{\textsuperscript{1}Institute of Astronomy and Astrophysics, Universit\'e Libre de Bruxelles, CP 226, Boulevard du Triomphe, B-1050 Brussels, Belgium.}
%\affilTwo{\textsuperscript{2}Department of Q, University Z, Place Pincode, Country.}

%%escape two column mode for title, affiliation and abstract
%%by giving \twocolumn command as shown

\twocolumn[{

\maketitle

%%include \corres to print the corresponding author Email id
\corres{nchamel@ulb.ac.be}

%%include \msinfo for
%%manuscript information such as
%%received, revised and accepted dates
%%
%\msinfo{1 January 2015}{1 January 2015}{1 January 2015}

%%abstract
\begin{abstract}
Neutron stars, the compact stellar remnants of core-collapse supernova explosions, are unique cosmic laboratories for exploring novel phases of matter under extreme conditions. In particular,  the occurrence of superfluidity and superconductivity in neutron stars will be 
briefly reviewed.
\end{abstract}

%%insert keywords separated by 3 hyphens using \keywords{words}

}]
%close the twocolumn escape here

%%include \doinum{number}for the DOI number in the header
%%include \volnum{number} for the volume number in the header
%%include \year{yyyy} for  year of publication in the header
%%include \pgrange{num--num} page range of article in the header
%%include \artcitid{num} for the article citation id
%%include \lp to print last page of the article
%%include \setcounter{page}{pagenum} for the exact starting page of the article

\doinum{10.1007/s12036-017-9470-9}
%\artcitid{43}
%\volnum{38}
\year{2017}
\pgrange{1}
\setcounter{page}{1}
\lp{14}

\section{Introduction}

Formed in the furnace of gravitational core-collapse supernova explosions of stars with a mass between 8 and 
 10 times that of the Sun~\cite{deshpande1995}, neutron stars contain matter crushed at densities exceeding  
that found inside the heaviest atomic nuclei (for a general review about neutron stars, see, e.g. Refs.~\cite{srinivasan1997,haensel2007}). 
A proto neutron star is initially fully fluid 
with a mass of about one or two solar masses, a radius of about 50 km and 
internal temperatures of the order $10^{11}-10^{12}$~K (for a review about neutron-star formation, see, e.g. 
Ref.~\cite{prakash2001}). About one minute later, the proto-neutron star becomes transparent to neutrinos that are copiously 
produced in its interior, thus rapidly cools down and shrinks into an ordinary neutron star. After a few months, 
the surface of the star - possibly surrounded by a very thin atmospheric plasma layer of light elements - still remains 
liquid. However, the layers beneath crystallize thus forming a solid crust~\cite{lrr}. At this point, the core is much colder 
than the crust because of the cooling power of the escaping neutrinos. After several decades, the interior of the star reaches a 
thermal equilibrium with temperatures of about $10^{8}$~K (except for a thin outer heat-blanketing envelope). The last cooling 
stage takes place after about a hundred thousand years, when heat from the interior diffuses to the surface and is dissipated in 
the form of radiation (for a recent review about neutron-star cooling, see, e.g. Ref.~\cite{potekhin2015}). 

With typical temperatures of order $10^7$~K, the highly degenerate matter in neutron stars is expected to become cold enough for the appearance of superfluids 
and superconductors -- frictionless quantum liquids respectively electrically neutral and charged~\cite{legett2006} -- made of neutrons
and protons, and more speculatively of other particles such as hyperons or quarks. If these phase transitions really occur, neutron stars would not only be
the largest superfluid and superconducting systems known in the Universe~\cite{srinivasan1997,lrr,sauls1989,sedrakian2006,page2014,graber2017}, but also 
the hottest ones with critical temperatures of 
the order of $10^{10}$~K as compared to a mere 203~K for the world record achieved in 2014 in terrestrial laboratories 
and consisting of hydrogen-sulphide compound under high pressure~\cite{droz2015}. 

After describing the main properties of terrestrial superfluids and superconductors, an overview of the theoretical developments 
in the modelling of superfluid and superconducting neutron stars will be given. Finally, the possible observational manifestations 
of these phases will be briefly discussed.

\section{Terrestrial superfluids and superconductors}

\subsection{Historical milestones}
\label{sec:history}

Superconductivity and superfluidity were known long before the discovery of pulsars in August 1967. Heike Kamerlingh Onnes and his collaborators were the 
first to liquefy helium in 1908, thus allowing them to explore the properties of materials at lower temperatures than could be reached before. On April 
8th, 1911, they observed that the electric resistance of mercury dropped to almost zero at temperature $T_c\simeq 4.2$~K
(for an historical account of this discovery, see e.g., Ref.~\cite{vandelft2010}). Two years later, 
lead and tin were found to be also superconducting. In 1914, Onnes showed that superconductivity is destroyed if the magnetic field exceeds some 
critical value. He later designed an experiment to measure the decay time of a magnetically induced electric current in a superconducting lead ring, 
and did not notice any change after an hour. Superconducting currents can actually be sustained for more than hundred 
thousand years~\cite{file1963}. In 1932, Willem Keesom 
and Johannes Antonie Kok found that the heat capacity of tin exhibits a discontinuity as it becomes superconducting thus demonstrating that this phase transition is of 
second order~\cite{keekok1932}. One year later, Walther Meissner and Robert Ochsenfeld made the remarkable observation that when a superconducting material 
initially placed in a magnetic field is cooled below the critical temperature, the magnetic flux is expelled from the sample~\cite{meissner1933}. This 
showed that superconductivity represents a new thermodynamical equilibrium state of matter. In 1935, Lev Vasilievich Shubnikov~\cite{shubnikov1935a,shubnikov1935} at the Kharkov Institute 
of Science and Technology in Ukraine discovered that some so called "hard" or type II superconductors (as opposed to "soft" or type I superconductors) 
exhibit two critical fields, between which the magnetic flux partially penetrates the material. Various superconducting materials were discovered in the 
following decades.

During the 1930's, several research groups in Leiden, Toronto, Moscow, Oxford and Cambridge (United Kingdom), found that below $T_\lambda \simeq 2.17$~K, helium-4 
(referred to as helium II) does not behave like an ordinary liquid (for a review of the historical context, see e.g., Refs.~\cite{balibar2007,balibar2014}). 
In particular, helium II does not boil, as was actually first noticed by Kamerlingh Onnes and his collaborators the same day they discovered 
superconductivity~\cite{vandelft2010}. Helium II can flow without resistance through very narrow slits and capillaries, almost independently of the pressure 
drop. The term ``superfluid'' was coined by Pyotr Kapitsa in 1938 by analogy with superconductors~\cite{kapitsa1938}. Helium II also flows up over the sides 
of a beaker and drip off the bottom (for ordinary liquids, the so called Rollin film is clamped by viscosity). The existence of persistent currents in 
helium II was experimentally established at the end of the 1950's and the beginning of 1960's~\cite{reppy1964}. The analog of the Meissner-Ochsenfeld phenomenon, 
which was predicted by Fritz London, was first observed by George Hess and William Fairbank at Stanford in June 1967~\cite{hess1967}: the angular momentum of helium-4 in a slowly 
rotating container was found to be reduced as the liquid was cooled below the critical temperature $T_\lambda$. 

At the time the first observed pulsars were identified as neutron stars, several materials had thus been found to be superconducting, while helium-4 was the 
unique superfluid known. The superfluidity of helium-3 was established by Douglas Osheroff, Robert Richardson and David Lee in 1971~\cite{osheroff72}. 
No other superfluids were discovered during the next two decades until the production of ultracold dilute gases of bosonic atoms in 1995~\cite{anderson1995,ketterle1995}, 
and of fermionic atoms in 2003~\cite{regal2004}. The main properties of some known superfluids and superconductors are summarized in Table~\ref{tab1}. 

\begin{table}[htb]
%% use tabular font for a smaller size font
\tabularfont
\caption{Properties of various superfluid and superconducting systems in order of their critical temperature $T_c$. Adapted from Table 1.1 in Ref.~\cite{legett2006}.}
\label{tab1} 
\begin{tabular}{lcr}
\topline
System        & Density (cm$^{-3}$)  & $T_c$ (K)  \\\midline
% \noalign{\smallskip}\svhline\noalign{\smallskip}
Neutron stars & $\sim 10^{39}$ & $\sim 10^{10}$ \\  
Cuprates  & & \\
and other exotics &  $\sim 10^{21}$ & $1-165$ \\ 
Electrons & & \\
 in ordinary metals & $\sim 10^{23}$ & $1-25$\\ 
Helium-4 & $\sim 10^{22}$ & $2.17$ \\
Helium-3 & $\sim 10^{22}$ & $2.491 \times 10^{-3}$ \\
Fermi alkali gases & $\sim 10^{12}$ & $\sim 10^{-6}$ \\
Bose alkali gases & $\sim 10^{15}$ & $\sim 10^{-7}-10^{-5}$ \\
\hline
\end{tabular}
%%use \tablenotes{footnote} to get the table foot note
% \tablenotes{Sample table footnote}%%9/11
\end{table}

\subsection{Quantum liquids}
\label{sec:quantum_liquids}

Superconductivity and superfluidity are among the most spectacular macroscopic manifestations of quantum mechanics. 
Satyendra Nath Bose and Albert Einstein predicted in 1924-1925 that at low enough temperatures an ideal gas of bosons
condense into a macroscopic quantum state~\cite{bose1924,einstein1925}. The association between Bose-Einstein condensation (BEC) and superfluidity 
was first advanced by Fritz London~\cite{london1938}. The only known superfluid at the time was helium-4, which is a boson. 
The condensate can behave coherently on a very large scale and can thus flow without any resistance. It was a key idea for 
developing the microscopic theory of superfluidity and superconductivity. Soon afterwards, Laszlo Tisza~\cite{tisza1938} 
postulated that a superfluid such as He II contains two distinct dynamical components: the condensate, which carries no 
entropy, coexists with a normal viscous fluid. This model explained all phenomena observed at the time and predicted 
thermomechanical effects like ``temperature waves''. Although Landau~\cite{landau1941} incorrectly believed that superfluidity 
is not related to BEC, he developed the two-fluid model and showed in particular that the normal fluid consists of 
``quasiparticles'', which are not real particles but complex many-body motions. This two-fluid picture was later adapted 
to superconductors~\cite{gorter1955}.

According to the microscopic theory of superconductivity by John Bardeen, Leon Cooper and Robert Schrieffer (BCS) published in 1957~\cite{bcs}, the 
dynamical distorsions of the crystal lattice (phonons) in a solid can induce an attractive effective interaction between 
electrons of opposite spins. Roughly speaking, electrons can thus form pairs and undergo a BEC below some critical temperature. 
A superconductor can thus be viewed as a charged superfluid. This picture however should not be taken too far. Indeed, electron pairs 
are very loosely bound and overlap. Their size $\xi\sim \hbar v_F/(k_B T_c)$ (usually referred to as the coherence length), with 
$v_F$ the Fermi velocity, $k_B$ Boltzmann's constant, and $T_c$ the critical temperature, is typically much larger than the lattice 
spacing. Moreover, electron pairs disappear at temperatures $T>T_c$. The BEC and the BCS transition are now understood as two different 
limits of the same phenomenon. The pairing mechanism suggested that fermionic atoms could also 
become superfluid, as was later confirmed by the discovery of superfluid helium-3. Since 2003, various other fermionic superfluids 
have been found, as mentioned in the previous section. 

As first discussed by Lars Onsager~\cite{onsager1949} and Richard Feynman~\cite{feynman1955}, the quantum nature of a superfluid is embedded 
in the quantization of the flow 
\begin{equation}
\label{eq.sect.super.dyn.vortices}
\oint \pmb{p} \cdot \pmb{d \ell} = N h \, ,
\end{equation}
where $\pmb{p}$ is the momentum per superfluid particle, $h$ denotes Planck's constant, $N$ is any integer, and the integral 
is taken over any closed path. It can be immediately recognized that this condition is the Bohr-Sommerfeld quantization 
rule. The flow quantization follows from the fact that a superfluid is a macroscopic quantum system whose momentum is thus 
given by $p=h/\lambda$, where $\lambda$ is the de Broglie wavelength. Requiring the length of any closed path to be an 
integral multiple of the de Broglie wavelength leads to Eq.~(\ref{eq.sect.super.dyn.vortices}). The physical origin of 
this condition has been usually obscured by the introduction of the ``superfluid velocity'' $\pmb{V_s}=\pmb{p}/m$, where $m$ is 
the mass of the superfluid particles. 

In a rotating superfluid, the flow quantization condition~(\ref{eq.sect.super.dyn.vortices}) leads to the appearence of $N$ quantised vortices. 
In a region free of vortices, the superflow is characterised by the irrotationality condition 
\begin{equation}\label{eq.sect.super.dyn.irrot}
\pmb{\nabla} \times \pmb{p} = 0\, .
\end{equation}
Inside a vortex, the superfluidity is destroyed. Because superfluid vortices are essentially of quantum nature, their internal structure cannot 
be described by a purely hydrodynamic approach. However, vortices can be approximately treated as structureless topological defects at length 
scales much larger than the vortex core size. As shown by Vladimir Konstantinovich Tkachenko~\cite{tkachenko1966}, quantized
vortices tend to arrange themselves on a regular triangular array, with a spacing given by 
\begin{equation}
\label{eq:d_upsilon}
d_\upsilon=\sqrt{\frac{h}{\sqrt{3} m \Omega}}\, ,
\end{equation}
where $\Omega$ is the angular frequency. Vortex arrays have been observed in superfluid helium~\cite{yarmchuk1979} 
and more recently in atomic Bose-Einstein condensates~\cite{abo2001,zwierlein2005}. 
At length scales much larger than the intervortex spacing $d_\upsilon$, the superfluid flow 
mimics rigid body rotation such that 
\begin{equation}
\label{eq.sect.super.dyn.irrot.vortices2}
\pmb{\nabla} \times \pmb{p} = m n_\upsilon \pmb{\kappa} \, ,
\end{equation}
where $n_\upsilon$ is the surface density of vortices given by
\begin{equation}
\label{eq:n_upsilon}
n_\upsilon = \frac{m \Omega}{\pi \hbar} \, ,
\end{equation}
and the vector $\pmb{\kappa}$, whose norm is equal to $h/m$, is aligned with the average angular velocity. Landau's original 
two-fluid model was further improved in the 1960's by Hall and Vinen~\cite{hall1956,hall1960}, and independently by Bekarevich and 
Khalatnikov~\cite{bekarevich1961} to account for the presence of quantized vortices within a coarse-grained average hydrodynamic 
description. 

The quantization condition~(\ref{eq.sect.super.dyn.vortices}) also applies to superconductors. But in this case, 
the momentum (in CGS units) is given by $\pmb{p}\equiv m\, \pmb{v}+(q/c)\pmb{A}$, where $m$, $q$, and $v$ are the mass, electric 
charge and velocity of superconducting particles respectively, and $\pmb{A}$ is the electromagnetic potential
vector. Introducing the density $n$ of superconducting particles and the ``supercurrent'' $\pmb{\cal J}=nq\pmb{v}$, 
the situation $N=0$ as described by Eq.~(\ref{eq.sect.super.dyn.irrot}) leads to the London equation
\begin{equation}
\label{eq.sect.super.dyn.London}
\pmb{\nabla} \times \pmb{\cal J} =  - \frac{c}{4\pi \lambda_L^2}\pmb{B} 
\end{equation}
where $\pmb{B} =\pmb{\nabla}\times \pmb{A}$ is the magnetic field induction, and $\lambda_L=\sqrt{m c^2/(4\pi n q^2)}$ is the London 
penetration depth. Situations with $N>0$ are encountered in 
type~II superconductors for which $\lambda_L\gtrsim \xi$. 
Considering a closed contour outside a sample of such a superconductor for which ${\cal J}=0$ and integrating the
momentum $\pmb{p}$ along this contour, leads to the quantization of the total magnetic flux $\Phi$ into fluxoids (also referred to as 
flux tubes or fluxons)
\begin{equation}
\label{eq:Phi}
\Phi=\oint \pmb{A} \cdot \pmb{d\ell} = N \Phi_0\, ,
\end{equation}
where $\Phi_0=h c/|q|$ is the flux quantum. The magnetic flux quantization, first envisioned by London, was experimentally confirmed in 
1961 by Bascom Deaver and William Fairbank at Stanford University~\cite{deaver1961}, and independently by Robert Doll and Martin N\"abauer 
at the Low Temperature institute in Hersching~\cite{doll1961}. As predicted by Aleksei Abrikosov~\cite{abrikosov1957}, these fluxoids 
tend to arrange themselves into a triangular lattice with a spacing given by
\begin{equation}
\label{eq:d_Phi}
d_\Phi=\sqrt{\frac{2 h c}{\sqrt{3} |q| B}}\, .
\end{equation}
Averaging at length scales much larger than $d_\upsilon$, the surface
density of fluxoids is given by
\begin{equation}
\label{eq:n_Phi}
n_\Phi = \frac{B}{\Phi_0}= \frac{|q| B}{h c} \, ,
\end{equation}
%NEW
where $B$ denotes here the average magnetic field strength. 
The size of a fluxoid (within which the superconductivity is destroyed) is of the order of the coherence length $\xi$. The magnetic field carried 
by a fluxoid extends over a larger distance of the order of the London penetration length $\lambda_L$. The nucleation of a single fluxoid thus occurs 
at a critical field $H_{c1}\sim \Phi_0/(\pi\lambda_L^2)$, and superconductivity is destroyed at the critical field $H_{c2}\sim \Phi_0/(\pi\xi^2)$ at 
which point the cores of the fluxoids touch. 

\section{Superstars}

\subsection{Prelude: internal constitution of a neutron star}

A few meters below the surface of a neutron star, matter is so compressed by the tremendous gravitational pressure that atomic nuclei, 
which are supposedly arranged on a regular crystal lattice, are fully ionized and thus coexist with a quantum gas of electrons. With increasing 
depth, nuclei become progressively more neutron-rich. Only in the first few hundred metres below the surface can the composition be 
completely determined by experimentally measured masses of atomic nuclei~\cite{wolf2013}. In the deeper layers recourse must be made 
to theoretical models (see, e.g., Refs.~\cite{pearson2011,hempel2013,chamel2015c,bcpm2015,utama2016,chamel2017}). At densities of a few $10^{11}$ 
g~cm$^{-3}$, neutrons start to ``drip'' out of nuclei (see, e.g., Ref.~\cite{cfzh2015} for a recent discussion). This marks the 
transition to the inner crust, an inhomogeneous assembly of neutron-proton clusters immersed in an ocean of unbound neutrons and 
highly degenerate electrons. 
According to various calculations, the crust dissolves into a 
uniform mixture of neutrons, protons and electrons when the density reaches about half the density 
$\sim 2.7\times 10^{14}$ g~cm$^{-3}$ 
found inside heavy atomic nuclei (see, e. g. Ref.~\cite{lrr} for a review about neutron-star crusts). Near the crust-core interface, nuclear clusters with very unusual shapes such as elongated rods or slabs may exist 
(see, e.g., Section 3.3 of Ref.~\cite{lrr}, see also Ref.~\cite{watanabe2012}). These so-called ``nuclear pastas'' could account for half of the crustal mass, and play a 
crucial role for the dynamical evolution of the star and its cooling~\cite{pons2013,horowitz2015}. The composition of the innermost part 
of neutron-star cores remains highly uncertain: apart from nucleons and leptons, it may also contain hyperons, meson condensates, and 
deconfined quarks (see, e.g. Ref.~\cite{haensel2007}; see also Refs.~\cite{sedrakian2010,chatterjee2016}). 

\subsection{Superfluid and superconducting phase transitions in dense matter}

Only one year after the publication of the BCS theory of superconductivity, 
Bogoliubov was the first to consider the possibility of superfluid nuclear matter~\cite{bogoliubov1958}. In 1959, 
Arkady Migdal~\cite{migdal1959} speculated that the interior of a neutron star might contain 
a neutron superfluid, and 
its critical temperature was estimated by Vitaly Ginzburg and David Abramovich Kirzhnits in 1964 using the BCS theory~\cite{ginzburg1964}. 
Proton superconductivity in neutron stars was studied by Richard Wolf in 1966~\cite{wolf1966}. The possibility of anisotropic 
neutron superfluidity was explored by Hoffberg, Glassgold, Richardson, and Ruderman~\cite{hoffberg1970}, and 
independently by Tamagaki in 1970~\cite{tamagaki1970}.

Neutrons and protons are fermions, and due to the Pauli exclusion principle, they generally tend to avoid themselves. This individualistic behaviour, 
together with the strong repulsive nucleon-nucleon interaction at short distance, provide the necessary pressure to counterbalance 
the huge gravitational pull in a neutron star, thereby preventing it from collapsing. However at low enough temperatures, nucleons 
may form pairs~\cite{50bcs} similarly as electrons in ordinary superconductors as described by the BCS theory\footnote{The high temperatures $\sim 10^7$~K prevailing 
in neutron stars interiors prevent the formation of electron pairs recalling that the highest critical temperatures of terrestrial 
superconductors do not exceed $\sim 200$~K. In particular, iron expected to be present in the outermost layers of a neutron star was found to 
be superconducting in 2001, but with a critical temperature $T_c\simeq 2$~K~\cite{shimizu2001}. See also Ref.~\cite{ginzburg1969}.}. 
These bosonic pairs can therefore condense, analogous to superfluid helium-3. 
While helium-3 becomes a superfluid only below 1 mK, nuclear superfluidity could be sustainable even at a temperature of several billions degrees in 
a neutron star due to the enormous pressure involved. The nuclear pairing phenomenon is also supported by the properties of atomic nuclei (see, e.g. 
Ref.~\cite{dean2003}). 

Because the nuclear interactions are spin dependent and include non-central tensor components (angular momentum dependent), different 
kinds of nucleon-nucleon pairs could form at low enough temperatures. The most attractive pairing channels\footnote{A given channel is denoted by $^{2S+1}L_J$, 
where $J$ is the total angular momentum, $L$ is the orbital angular momentum, and $S$ the spin of nucleon pair.} are $^1$S$_0$ at low densities 
and the coupled $^3$PF$_2$ channel at higher densities~\cite{gezerlis2014}. In principle, different types of pairs may coexist. However, one or 
the other are usually found to be energetically favored~\cite{lombardo2001}. Let us mention that nucleons may also form quartets such as 
$\alpha$-particles, which can themselves condense at low enough temperatures (see, e.g. Ref.~\cite{schuck2014}). 
Most microscopic calculations have been carried out in pure neutron matter using diagrammatic, variational, and more recently Monte Carlo methods (see, e.g., 
Refs.~\cite{gezerlis2014,lombardo2001} for a review). At concentrations below $\sim 0.16$ fm$^{-3}$, as encountered in the inner crust and in the outer core of a 
neutron star, neutrons are expected to become superfluid by forming $^{1}$S$_0$ pairs, with critical temperatures of about $10^{10}$~K at most (see, e.g. 
Refs.~\cite{gezerlis2014,cao2006,maurizio2014,ding2016}). At neutron concentrations above $\sim 0.16$ fm$^{-3}$, pairing in the coupled $^3$PF$_2$ channel 
becomes favored but the maximum critical temperature remains very uncertain, predictions ranging from $\sim 10^8$~K to $\sim 10^9$~K (see, e.g. 
Refs.\cite{maurizio2014,ding2016,baldo1998,dong2013}). This lack of knowledge of neutron superfluid properties mainly stems from the highly nonlinear character 
of the pairing phenomenon, as well as from the fact that the nuclear interactions are not known from first principles (see, e.g., Ref.~\cite{machleidt2017} for a 
recent review). 

Another complication arises from the fact that neutron stars are not only made of neutrons. The presence of nuclear clusters in the crust of a neutron star may 
change substantially the neutron superfluid properties. Unfortunately, microscopic calculations of inhomogeneous crustal matter employing realistic nuclear 
interactions are not feasible. State-of-the-art calculations are based on the nuclear energy density functional theory, which allows for a consistent and unified 
description of atomic nuclei, infinite homogeneous nuclear matter and neutron stars (see, e.g. Ref.~\cite{chamel2013} and references therein). The main limitation 
of this approach is that the \emph{exact} form of the energy density functional is not known. In practice, phenomenological functionals fitted to selected nuclear 
data must therefore be employed. The superfluid in 
neutron-star crusts, which bears similarities with terrestrial multiband superconductors, was first studied within the band theory of solids in Ref.~\cite{chamel2010}. 
However, this approach is computationally very expensive, and has been so far limited to the deepest layers of the crust. For this reason, most calculations of 
neutron superfluidity in neutron-star crusts (see, e.g. Ref.~\cite{margueron2012}) have been performed using an approximation introduced by Eugene Wigner and 
Frederick Seitz in 1933 in the context of electrons in metals~\cite{wigner1933}: the Wigner-Seitz or Voronoi cell of the lattice (a truncated octahedron in case 
of a body-centred cubic lattice) is replaced by a sphere of equal volume. However, this approximation can only be reliably applied in the shallowest region of the 
crust due to the appearance of spurious shell effects~\cite{chamel2007}. Such calculations have shown that the phase diagram of the neutron superfluid in the crust 
is more complicated than that in pure neutron matter; in particular, the formation of neutron pairs can be enhanced with increasing temperature (see, e.g., 
Refs.~\cite{margueron2012b,pastore2013,pastore2015}). Microscopic calculations in pure neutron matter at densities above the crust-core boundary  
are not directly applicable to neutron stars due to the presence of protons, leptons, and possibly other particles in neutron-star cores. Few microscopic calculations 
have been 
performed so far in beta-stable matter (see, e.g., Ref.~\cite{zhou2004}). Because the proton concentration in the outer core of a neutron star 
is very low, protons are expected to become superconducting in the $^1$S$_0$ channel. However, the corresponding critical temperatures are very poorly known due 
to the strong influence of the surrounding neutrons~\cite{baldo2007}. Neutron-proton pairing could also in principle occur, but is usually disfavored by the 
very low proton content of neutron stars (see, e.g., Ref.~\cite{stein2014}). Other more speculative possibilities include hyperon-hyperon and hyperon-nucleon 
pairing (see, e.g. Ref.~\cite{chatterjee2016} and references therein). The core of a neutron star might also contain quarks in various color superconducting 
phases~\cite{alford2008}. 

According to cooling simulations, the temperature in a neutron star is predicted to drop below the estimated critical temperatures of nuclear superfluid phases 
after $\sim 10-10^2$ years. 
The interior of a neutron star is thus thought to contain at least three different kinds of superfluids and superconductors~\cite{page2014}: (i) an 
isotropic neutron superfluid (with $^1$S$_0$ pairing) permeating the inner region of the crust and the outer core, (ii) an anisotropic neutron superfluid (with 
$^3$PF$_2$ pairing) in the outer core, and (iii) an isotropic proton superconductor (with $^1$S$_0$ pairing) in the outer core. 
The neutron superfluids in the crust and in the outer core are not expected to be separated by a normal region. 

\subsection{Role of a high magnetic field}

Most neutron stars that have been discovered so far are radio pulsars with typical surface magnetic fields of order $10^{12}$~G (as compared to $\sim 10^{-1}$~G 
for the Earth's magnetic field), but various other kinds of neutron stars have been revealed with the development of the X-ray and gamma-ray astronomy~\cite{harding2013}. 
In particular, a small class of very highly magnetised neutron stars thus dubbed 
\emph{magnetars} by Robert Duncan \& Christopher Thompson in 1992~\cite{thompson1992} (see e.g. Ref.~\cite{woods2006} for a review) have been identified in the form 
of soft-gamma ray repeaters (SGRs) and anomalous x-ray pulsars (AXPs). Tremendous magnetic fields up to about $2\times 10^{15}$~G 
have been measured at the surface of these stars from both spin-down and spectroscopic studies~\cite{mcgill2014,tien2013,hong2014}, and 
various observations suggest the existence of even higher internal fields~\cite{ste05,kam07,ws07,sa07,vie07,rea10,maki14}. Although 
only 23 such stars are currently known~\cite{mcgill2014}, recent observations indicate that ordinary pulsars can also be endowed with 
very high magnetic fields of order $10^{14}$~G~\cite{ka11}. According to numerical simulations, neutron stars may potentially be endowed with 
internal magnetic fields as high as $10^{18}$~G  (see, e.g. Refs.~\cite{pili14,chatterjee2015} and references therein).

The presence of a high magnetic field in the interior of a neutron star may have a large impact on the superfluid and superconducting 
phase transitions. Proton superconductivity is predicted to disappear at a critical field of order $10^{16}-10^{17}$~G~\cite{baym1969a}. 
Because spins tend to be aligned in a magnetic field, the formation of neutron pairs in the $^1$S$_0$ channel is disfavored in a highly 
magnetized environment, as briefly mentioned by Kirzhnits in 1970~\cite{kirzhnits1970}. It has been recently shown that $^1$S$_0$ pairing in pure 
neutron matter is destroyed if the magnetic field 
strength exceeds $\sim 10^{17}$~G~\cite{stein2016}. 
Moreover, the magnetic field may also shift the onset of the neutron-drip transition in dense matter  to higher or lower densities due to Landau quantization 
of electron motion thus changing the spatial extent of the superfluid region in magnetar crusts~\cite{csmmpv2015,fantina2016,basilico2015,chamel2016b}. 

\subsection{Dynamics of superfluid and superconducting neutron stars}
\label{sec:superstars_hydro}

The minimal model of superfluid neutron stars consists of at least two distinct interpenetrating dynamical components~\cite{baym1969}: 
(i) a plasma of electrically charged particles (electrons, nuclei in the crust and protons in the core) that are essentially locked together 
by the interior magnetic field, and (ii) a neutron superfluid. Whether protons in the core are superconducting or not, they co-move with the 
other electrically charged particles (see, e.g. Ref.~\cite{sauls1989}). 

The traditional heuristic approach to superfluid hydrodynamics blurring the distinction 
between velocity and momentum makes it difficult to adapt and extend Landau's original two-fluid model to the relativistic context, as required for 
a realistic description of neutron stars (see, e.g. Ref.~\cite{carter1994}). 
In particular, in superfluid mixtures such as helium-3 and helium-4~\cite{andreev1976}, or neutrons and protons 
in the core of neutron stars~\cite{sedrakyan1980,vardanyan1981}, the different superfluids are generally mutually coupled by entrainment effects whereby the true 
velocity $\pmb{v_{\rm X}}$ and 
the momentum $\pmb{p_{\rm X}}$ of a fluid X are not aligned: 
\begin{equation}
\pmb{p_{\rm X}}=\sum_{\rm Y} \mathcal{K}^{\rm XY} \pmb{v_{\rm Y}}\, , 
\end{equation}
where $\mathcal{K}^{\rm XY}$ is a symmetric matrix determined by the interactions between the constituent particles. 
In the two-fluid model, entrainment can be equivalently formulated  
in terms of ``effective masses''. Considering the neutron-proton mixture in the core of neutron stars, the neutron momentum can thus 
be expressed as $\pmb{p_n}=m_n^\star \pmb{v_n}$ in the proton rest frame ($\pmb{v_p}=0$), with $m_n^\star=\mathcal{K}^{nn}$. Alternatively, 
a different kind of effective mass can be introduced, namely $m_n^\sharp=\mathcal{K}^{nn}-\mathcal{K}^{np}\mathcal{K}^{pn}/\mathcal{K}^{pp}$, such 
that $\pmb{p_n}=m_n^\sharp \pmb{v_n}$ in the proton momentum rest frame ($\pmb{p_p}=0$). 
These effectives masses should not be confused with those introduced in microscopic 
many-body theories (see, e.g. Ref.~\cite{chamel2006}). 
Because of the strong interactions between neutrons and protons, entrainment effects in 
neutron-star cores cannot be ignored (see, e.g. Refs.~\cite{chamel2006,gusakov2005,chamel2008,kheto2014,sourie2016} for recent estimates). 
In the neutron-rich core of neutron stars, we typically have $m_n^\star \sim m_n^\sharp \sim m_n$, and $m_p^\star \sim m_p^\sharp \sim (0.5-1) m_p$, 
where $m_n$ and $m_p$ denote the ``bare'' neutron and proton masses respectively. As shown by Brandon 
Carter in 1975~\cite{carter1975}, at the global 
scale of the star, general relativity induces additional couplings between the fluids due to Lense-Thirring effects, which tend to counteract entrainment. 
As recently found in Ref.~\cite{sourie2017}, frame-dragging effects can be as important as entrainment. 

An elegant variational formalism to derive the hydrodynamic equations of any relativistic (super)fluid mixtures was developed by Carter and 
collaborators 
(see, e.g., Refs.~\cite{carter1989,carter2001,gourgoulhon2006,andersson2007}). 
This formalism relies on an action principle 
in which the basic variables are the number densities and currents of the different fluids. 
The equations of motion can be derived by considering variations of the fluid particle trajectories. 
Dissipative processes (e.g. viscosity in non-superfluid constituents, 
superfluid vortex drag, 
ordinary resistivity 
between non-superfluid constituents, nuclear reactions) can be treated within the same framework. 
The convective formalism developed by Carter was later adapted to the comparatively more intrincate Newtonian theory within a 4-dimensionally covariant 
framework~\cite{carter2004,carter2005,carter2005b} (see, e.g. Refs.~\cite{prix2004,prix2005,andersson2006} and references therein for a review of other 
approaches using a 3+1 spacetime decomposition). This fully covariant approach not only facilitates the comparison with the relativistic theory (see, e.g., 
Refs.~\cite{carter2006,chamel2008}), but more importantly lead to the discovery of new conservation laws in superfluid systems such as the conservation of 
generalised helicy currents. 

As pointed out by Ginzburg and Kirzhnits in 1964~\cite{ginzburg1964}, the interior of a rotating neutron star is expected to be threaded by a very large number 
of neutron superfluid vortices (for a discussion of the vortex structure in $^1$S$_0$ and $^3$PF$_2$ neutron superfluids, see, e.g. Ref.~\cite{sauls1989}). 
Introducing the spin period $P$ in units of 10 ms, $P_{10}\equiv P/(10~{\rm ms})$, the surface density of vortices 
(\ref{eq:n_upsilon}) is of the order
\begin{equation}
 n_\upsilon \sim 6\times 10^5\, P_{10}^{-1} \ {\rm cm}^{-2} \, .
\end{equation}
The average intervortex spacing (\ref{eq:d_upsilon})
\begin{equation}
 d_\upsilon \sim n_\upsilon^{-1/2} \sim  10^{-3}\, \sqrt{P_{10}} \ {\rm cm} \, ,
\end{equation}
is much larger than the size of the vortex core (see, e.g. Ref.~\cite{yu2003}). 
Neutron superfluid vortices can pin to nuclear inhomogeneities in the crust. However, the pinning strength remains uncertain (see, e.g. Ref.~\cite{wlazlowski2016}
and references therein; see also Section 8.3.5 of Ref.~\cite{lrr}). 
Protons in the core of a neutron star are expected to become superconducting at
low enough temperatures. Contrary to superfluid neutrons, superconducting protons do not form vortices. 
As shown by Baym, Pethick, and Pines in 1969~\cite{baym1969a,baym1969b}, the expulsion of the magnetic flux accompanying the transition 
takes place on a very long time scale 
$\sim 10^6$ years due to the very high electrical conductivity of the dense stellar matter. The superconducting transition 
thus occurs at constant magnetic flux. The proton superconductor is usually thought to be of type II~\cite{baym1969a} (but see also 
Refs.~\cite{charbonneau2007,alford2008} and references therein), 
in which case, the magnetic flux penetrates the neutron star core by forming fluxoids, with a surface density~(\ref{eq:n_Phi}) of order
\begin{equation}
 n_\Phi \sim 5\times 10^{18}\, B_{12} \ {\rm cm}^{-2} \, ,
\end{equation}
where the magnetic field strength $B$ is expressed as $B_{12}\equiv B/(10^{12}~{\rm G})$. This surface density corresponds to a spacing 
\begin{equation}
 d_\Phi \sim n_\Phi^{-1/2} \sim 5\times 10^{-10}\, \sqrt{B_{12}}^{-1} \ {\rm cm} \, .
\end{equation}
Since the magnetic flux is frozen in the stellar core, fluxoids can form even if the magnetic field is lower than the critical field 
$H_{c1}\sim 10^{15}$~G~\cite{baym1969a}. Proton superconductivity is destroyed at the higher critical field $H_{c2}\sim 10^{16}$~G~\cite{baym1969a}. 
Due to entrainment effects, neutron superfluid vortices carry a magnetic flux as well, given by~\cite{sedrakyan1980,alpar1984}
\begin{equation}
 \Phi=\Phi_0\left(\frac{m_p^\sharp}{m_p}-1\right)\, ,
\end{equation}
where $\Phi_0=h c/(2 e)$. 
Electrons scattering off the 
magnetic field of the vortex lines leads to a strong frictional coupling between the core neutron superfluid and the electrically charged 
particles~\cite{alpar1984}. Neutron superfluid vortices could also interact with proton fluxoids~\cite{sauls1989,muslimov1985,mendell1991,chau1992}, 
and this may have important implications for the evolution of the star~\cite{srinivasan1997,sauls1989,srinivasan1990,ruderman1995,ruderman1998,bhattacharya2002}. 
For typical neutron star parameters ($P=10$~ms, $B=10^{12}$~G, radius $R=10$~km), the numbers of neutron superfluid vortices 
and proton fluxoids are of order $n_\upsilon \pi R^2\sim 10^{18}$ and $n_\Phi \pi R^2\sim 10^{30}$, respectively. Such large numbers justify a smooth-averaged 
hydrodynamical description of neutron stars. However, this averaging still requires the understanding of the underlying vortex dynamics 
(see, e.g. Ref.~\cite{graber2017}). A more elaborate treatment accounting for the macroscopic anisotropy induced by the underlying presence of vortices 
and/or flux tubes was developed by Carter based on a Kalb-Ramond type formulation~\cite{carter2000} (see also Ref.~\cite{gusakov2016} and references therein). 
In recent years, simulations of large collections ($\sim 10^2 - 10^4$) of vortices have been carried out, thus providing some 
insight on collective behaviors, such as vortex avalanches (see, e.g., Ref.~\cite{warszawski2013}). However, these simulations have been restricted so far 
to Bose condensates. The extent to which the results can be extrapolated to neutron stars remains to be determined. 
Such large-scale simulations also require microscopic parameters determined by the local dynamics of individual vortices (see e.g. Ref.~\cite{bulgac2013}).

The variational formulation of multifluid hydrodynamics was extended for studying the magnetoelastohydrodynamics of neutron star crusts, allowing for 
a consistent treatment of the elasticity of the crust, superfluidity and the presence of a strong magnetic field, both within the Newtonian 
theory~\cite{carter2006,carter2006b} and in the fully relativistic context~\cite{carter2006c}. In particular, these formulations can account for 
the entrainment of the neutron superfluid by the crustal lattice~\cite{carter2006d}, a non-dissipative effect arising from Bragg scattering of 
unbound neutrons first studied in Refs.~\cite{carter2005c,chamel2005,chamel2006b} using the band theory of solids. More recent systematic calculations 
based on a more realistic description of the crust have confirmed that these entrainment effects can be very strong~\cite{chamel2012}. These results 
are at variance with those obtained from hydrodynamical studies~\cite{epstein1988,sedrakian1996,magierski2004,magierski2004b,magierski2004c,martin2016}. 
However, as discussed in Ref.~\cite{martin2016}, these approaches are only valid if the neutron superfluid coherence length is much smaller than the 
typical size of the spatial inhomogeneities, a condition that is usually not fulfilled in most region of the inner crust. The neglect of neutron pairing 
in the quantum calculations of  Ref.~\cite{chamel2012} has been recently questioned~\cite{gezerlis2014,martin2016}. Although detailed numerical calculations 
are still lacking, the analytical study of Ref.~\cite{carter2005d} suggests that neutron pairing is unlikely to have a large impact on the entrainment coupling.

\section{Observational manifestations}

\subsection{Pulsar frequency glitches}

Pulsars are neutron stars spinning very rapidly with extremely stable periods $P$ ranging from milliseconds to seconds, with delays 
$\dot{P}\equiv dP/dt$ that in some cases do not do not exceed $10^{-21}$, as compared to $10^{-18}$ for the most accurate atomic 
clocks~\cite{hinkley2013}. Nevertheless, irregularities have been detected in long-term pulsar timing observations (see, e.g., Ref.~\cite{lyne1995}). 
In particular, some pulsars have been found to suddenly spin up. These ``glitches'' in their rotational frequency $\Omega$, ranging from 
$\Delta\Omega\slash \Omega\sim 10^{-9}$ to $\sim 10^{-5}$, are generally followed by a long relaxation lasting from days to years, and sometimes 
accompanied by an abrupt change of the spin-down rate from $\vert\Delta \dot\Omega\slash\dot\Omega\vert\sim 10^{-6}$ up to $\sim 10^{-2}$. 
At the time of this writing, 482 glitches have been detected in 168 pulsars~\cite{jod12}. Since these phenomena have not been observed in 
any other celestial bodies, they must reflect specific properties of neutron stars (for a recent review, see, e.g. Ref.~\cite{haskell2015}). 
In particular, giant pulsar frequency glitches $\Delta\Omega\slash \Omega\sim 10^{-6}-10^{-5}$ as detected in the emblematic Vela pulsar are 
usually attributed to sudden transfers of angular momentum from a more rapidly rotating superfluid component to the rest of star whose rotation 
frequency is directly observed (for a short historical review of theoretical developments, see, e.g. Ref.~\cite{chamel2015b} and references therein). 
The role of superfluidity is corroborated by the very long relaxation times~\cite{baym1969a} and by experiments with superfluid helium~\cite{tsakadze1980}. 
The standard scenario of giant pulsar glitches is the following. The inner crust of a neutron star is permeated by a neutron superfluid that is weakly 
coupled to the electrically charged particles by mutual friction forces 
(in a seminal work, Alpar, Langer and Sauls~\cite{alpar1984} argued that the 
core neutron superfluid is strongly coupled to the core, and therefore does not participate to the glitch). 
The superfluid thus follows the 
spin-down of the star via the motion of vortices away from the rotation axis unless vortices are pinned to the crust~\cite{anderson1975}. In such a case, 
a lag between the superfluid and the rest of the star will build up, inducing a Magnus force acting on the vortices. At some point, the vortices will suddenly unpin, 
the superfluid will spin down and, by the conservation of angular momentum the crust will spin up. During the subsequent relaxation, vortices progressively 
repin until the next glitch~\cite{alpar1985}. This scenario is supported by the analysis of the glitch data, suggesting that the superfluid represents only 
a few per cent of the angular momentum reservoir of the star~\cite{alpar1993,datta1993,link1999}. On the other hand, this interpretation has been recently 
challenged by the 2007 glitch detected in PSR~J1119$-$6127, and by the 2010 glitch in PSR~B2334$+$61~\cite{yuan2010,alpar2011,akbal2015}. More importantly, 
it has been also shown that the neutron superfluid in the crust of a neutron star does not contain enough angular momentum to explain giant glitches due to 
the previously ignored effects of Bragg scattering~\cite{chamel2006c,andersson2012,chamel2013b,delsate2016}. This suggests that the core superfluid plays a 
more important role than previously thought~\cite{ho2015,pizzochero2017}. In particular, the core superfluid could be decoupled from the rest of the star due 
to the pinning of neutron vortices to proton fluxoids~\cite{ruderman1998,gugercinoglu2014}. So far, most global numerical simulations of pulsar glitches have 
been performed within the Newtonian theory (see, e.g. Refs.~\cite{larson2002,peralta2006,sidery2010,haskell2012}). However, a recent study shows that general relativity 
could significantly affect the dynamical evolution of neutron stars~\cite{sourie2017}. 

\subsection{Thermal relaxation of transiently accreting neutron stars during quiescence}

In a low-mass X-ray binary, a neutron star accretes matter from a companion star during several years or decades, driving the neutron-star crust 
out of its thermal equilibrium with the core. After the accretion stops, the heated crust relaxes towards equilibrium (see, e.g., Section 12.7 of Ref.~\cite{lrr}, 
see also Ref.~\cite{page2012}). The thermal relaxation has been already monitored in a few systems (see, e.g., Ref.~\cite{waterhouse2016} and references therein). 
The thermal relaxation time depends on the properties of the crust, especially the heat capacity. In turn, the onset of neutron superfluidity leads to a strong 
reduction of the heat capacity at temperatures $T\ll T_c$ thus delaying the thermal relaxation of the crust (see, e.g., Ref.~\cite{fortin2010}). If neutrons were 
not superfluid, they could store so much heat that the thermal relaxation would last longer than what is observed~\cite{shternin2007,brown2009}. On the other hand, 
the thermal relaxation of these systems is not completely understood. For instance, additional heat sources of unknown origin are needed in order to reproduce the 
observations~\cite{waterhouse2016,brown2009,degenaar2013,degenaar2014,turlione2015,degenaar2015,merritt2016}. These discrepancies may also originate from a lack 
of understanding of superfluid properties~\cite{turlione2015}. In particular, the low-energy collective excitations of the neutron superfluid were found to be 
strongly mixed with the vibrations of the crystal lattice, and this can change substantially the thermal properties of the crust~\cite{chamel2013c,chamel2016}. 

\subsection{Rapid cooling of Cassiopeia A}

Cassiopeia A is the remnant of a star that exploded 330 years ago at a distance of about 11000 light years from us. It owes its name to its location 
in the constellation Cassiopeia. The neutron star is not only the youngest known, thermally emitting, isolated neutron star in our Galaxy, but it is 
also the first isolated neutron star for which the cooling has been directly observed. Ten-year monitoring of this 
object 
seems to indicate that 
its temperature has 
has decreased 
by a few percent since its discovery in 1999~\cite{heinke2010} (but see also 
the analysis of Refs.~\cite{elshamouty2013,posselt2013} suggesting that the temperature decline 
is not statistically significant). 
If confirmed, this cooling rate would be substantially  
faster than that expected from nonsuperfluid neutron-star cooling theories. It is thought that the onset of neutron superfluidity 
opens a new channel for neutrino emission from the continuous breaking and formation of neutron pairs. This process, which is most effective for 
temperatures slightly below the critical temperature of the superfluid transition, enhances the cooling of the star during several decades. As a 
consequence, observations of Cassiopeia A put stringent constraints on the critical temperatures of the neutron superfluid and proton superconductor 
in neutron-star cores~\cite{page2011,shternin2011,ho2015b}. However, this interpretation has been questioned and alternative scenarios have 
been proposed~\cite{blaschke2013,negreiros2013,sedrakian2013,noda2013,bonanno2014,ouyed2015,sedrakian2016,taranto2016} (most of which still requiring superfluidity 
and/or superconductivity in neutron stars).

\subsection{Pulsar timing noise and rotational evolution} 

Apart from pulsar frequency glitches, superfluidity and superconductivity may leave their imprint on other timing irregularities. In particular, pulsar 
timing noise (see, e.g. Ref.~\cite{lyne1995}) could be the manifestation of superfluid turbulence although other mechanisms 
are likely to play a role (see, e.g. Ref.~\cite{melatos2014} and references therein). 
Interpreting the long-period ($\sim 100-1000$ days) oscillations in the timing residuals of some pulsars such as PSR~B1828$-$11 (see, e.g. Ref.~\cite{kerr2016}) 
as evidence of free precession, it has been argued that either the neutron superfluid does not coexist with the proton superconductor in the core of a neutron star, 
or the proton superconductor is type I so as to avoid pinning of neutron superfluid vortices to proton fluxoids~\cite{link2003,link2007}. However, this conclusion 
seems premature in view of the complexity of the neutron-star dynamics~\cite{alpar2005,glampedakis2009}. 
Alternatively, these oscillations might be related to the propagation of Tkachenko waves in the vortex lattice (see, e.g. Ref.\cite{haskell2011} and 
references therein). 
The presence of superfluids and superconductors in the interior of a neutron star may also be revealed from the long-term rotational evolution of 
pulsars by measuring the braking index $n=\Omega \ddot{\Omega}/\dot{\Omega}^2$. Deviations from the canonical value $n=3$ as predicted by a rotating magnetic dipole 
model 
in vacuum 
can be explained by the decoupling of the neutron superfluid in the core of a neutron star (due to pinning to proton fluxoids for instance)~\cite{alpar2006,ho2012}. 
However, a similar rotational evolution could be mimicked by other mechanisms without invoking superfluidity (see, e.g. Ref.~\cite{petri2016} for a recent review).

\subsection{Quasi-periodic oscillations in soft gamma-ray repeaters}

Quasi-periodic oscillations (QPOs) in the hard X-ray emission were detected in the tails of giant flares from SGR 1806-20, 
SGR 1900+14, and SGR 0526-66, with frequencies ranging from 18 Hz to 1800 Hz (see, e.g. Ref.~\cite{turolla2015} for a recent 
review). As anticipated by Duncan~\cite{duncan1998}, 
these QPOs are thought to be the signatures of global magneto-elastic seismic vibrations of the star. If this interpretation is 
confirmed, the analysis of these QPOs could thus provide valuable information on the interior of a neutron star. In particular, the
identification of the modes could potentially shed light on the existence of superfluid and superconducting phases 
(see, e.g. Ref.~\cite{gabler2013}). 

\section{Conclusion}

The existence of superfluid and superconducting phases in the dense matter constituting the interior of neutron stars 
has been corroborated both by theoretical developments and by astrophysical observations. In particular, neutron stars are expected to 
contain a $^1$S$_0$ neutron superfluid permeating the inner region of the crust and the outer core, a $^3$PF$_2$ neutron 
superfluid in the outer core, and a $^1$S$_0$ proton superconductor in the outer core. Still, many aspects of these 
phenomena need to be better understood. Due to the highly nonlinear character of the pairing mechanism giving rise to nuclear 
superfluidity and superconductivity, the associated critical temperatures remain very uncertain, especially for the $^3$PF$_2$ channel. 
The dynamics of these phase transitions as the star cools down, and the possible formation of topological defects need to be explored. 
Although the formalism for describing the relativistic smooth-averaged magnetoelastohydrodynamics of superfluid and superconducting
 systems already exists, modelling the global evolution of neutron stars in full general relativity still remains very challenging. To a large 
 extent, the difficulty lies in the many different scales involved, from the kilometer size of the star down to the size of individual 
  neutron vortices and proton fluxoids at the scale of tens or hundred fermis. 
  Studies of neutron-star dynamics using the Newtonian theory provide valuable qualitative insight, and should thus be pursued. 
The presence of other particles such as hyperons or 
  deconfined quarks in the inner core of neutron stars adds to the complexity. The occurrence of exotic superfluid and superconducting
   phases remains highly speculative due to the lack of knowledge of dense matter. On the other hand, astrophysical observations 
   offer a unique opportunity to probe the phase diagram of matter under extreme conditions that are inaccessible in terrestrial laboratories.

%%Use section* for acknowledgements
\section*{Acknowledgement}
This work has been supported by the Fonds de la Recherche Scientifique - FNRS (Belgium) under grant n$^\circ$~CDR J.0187.16, 
and by the European Cooperation in Science and Technology (COST) action MP1304 \emph{NewCompStar}. 

%%use \balance somewhere in the left column of the last page to balance the two columns in the end page

%%References section

\end{document}